\documentclass[%
 reprint,
 amsmath,amssymb,
 aps,
 prm,
]{revtex4-2}

\usepackage{easyReview}
\usepackage{graphicx}% Include figure files
\usepackage{dcolumn}% Align table columns on decimal point
\usepackage{bm}% bold math
\usepackage{hyperref}% add hypertext capabilities
\hypersetup{colorlinks=true, citecolor=blue, urlcolor=blue, linkcolor=blue}
\usepackage{mhchem}
\usepackage{multirow}
\usepackage{makecell}
\usepackage{numprint}
\usepackage{tabularx}
\usepackage[capitalize]{cleveref}
\usepackage[caption=false]{subfig}

\begin{document}

%\preprint{APS/123-QED}

\title{Structure of $180^\circ$ ferroelectric domain walls in $\mathrm{HfO_2}$ and $\mathrm{ZrO_2}$}% Force line breaks with \\

\author{Hongchu Du}
 \altaffiliation[ ]{Ernst Ruska-Centre for Microscopy and Spectroscopy with Electrons, Forschungszentrum Jülich GmbH, 52425 Jülich, Germany.}%Lines break automatically or can be forced with
  \altaffiliation[]{Central Facility for Electron Microscopy,
  	RWTH Aachen University,
  	Ahornstr. 55, 52074 Aachen, Germany.}%Lines break automatically or can be forced with \\
  \email{h.du@fz-juelich.de}

\date{\today}% It is always \today, today,

\begin{abstract}
We report two series of structures representing two types of $180^\circ$ ferroelectric  domain walls in $\mathrm{HfO_2}$ and $\mathrm{ZrO_2}$. We model the domain structures with different width by density functional theory calculations. The results reveal the dependency of domain wall energy and polarization on domain width. We show how the directions of polarization of ferroelectric domains can be determined from experimentally observed Hf atoms, despite the fact that the knowledge about the positions of O atoms is missing.  The two series of structures thus provide more thorough interpretation and hence deeper understanding of experimental observations by transmission electron microscopy.
%\begin{description}
%%Secondary publications and information retrieval purposes.
%%\end{description}
\end{abstract}

\maketitle
\section{introduction}
Ferroelectric domain walls are the interfaces between adjacent domains that have different polarization orientations.  Domain walls play an essential role in the nucleation and growth of ferroelectric domains through which the polarization switching occurs \cite{merzDomainFormation1954}. Electrical properties of domain walls may be dramatically different from those of the domains that are separated by the walls \cite{seidelConductionDW2009,mcconvilleMemristorDM2020}, showing great promise for application of domain walls as functional elements in future nanoelectronics. Due to the aforementioned fundamental and technological interest, understanding of the atomic structure of domain walls becomes an essential part of studies on ferroelectric materials \cite{seidelBook2016,tagantsevBook2010,natafDomainwall2020}.   

Thin films of pure and doped \ce{HfO2} and \ce{ZrO2}  were recently discovered to exhibit ferroelectric behavior \cite{bosckeFeHfO22011,mullerFerroHfZrO22012,nishimuraNondopedfilm2016}. The well-established technology for integration of these materials in silicon  promises application of ferroelectricity in information storage and computing devices \cite{chenReviewNVRAM2016,hwangFeRAM2019}. Enthusiasm for such devices hence has led to intensive studies on ferroelectric \ce{HfO2} and \ce{ZrO2} in the past decade \cite{parkRev2015}. Both \ce{HfO2} and \ce{ZrO2} have many crystallographic polymorphs. At atmospheric pressure, \ce{HfO2} exhibits a thermodynamic stable monoclinic (\textit{M}) phase at room temperature. The \textit{M} to a tetragonal (\textit{T}) phase and the \textit{T} to a fluorite-type cubic (\textit{C}) phase transformations occur at 2000 and 2800 K \cite{wangZirconiaHafniaSystemDTA2006}, respectively. \ce{ZrO2} exhibits similar phase transformations that  occur at 1470 and 2580 K \cite{navrotskyZrO22005}. All the \textit{M}, \textit{T}, and \textit{C} phases of \ce{HfO2} and \ce{ZrO2} are centrosymmetric and hence exhibit no spontaneous polarization.  Experimental results  of electron and neutron diffraction \cite{sangStructuralOrigins2015,xuSinglecrystal2021}, conventional TEM (CTEM) negative spherical aberration (Cs) imaging (NCSI) \cite{duMatter2021}, and bright-field scanning transmission electron microscopy (BF-STEM) imaging \cite{zhangEpitaxial2021} of materials in the form of thin films, nanocrystals, and single crystals  consistently confirmed that the ferroelectric behavior of \ce{HfO2} based materials is from a metastable polar orthorhombic  ($O_\mathrm{FE}$) phase having space group $Pbc2_1$, a different setting of $Pca2_1$ \cite{glazerSpaceGroupsSolid2013}.

High-angle annular dark-field (HAADF) STEM studies on Gd-doped polycrystalline films \cite{grimleyDomain2018} and Y-doped bulk singlcrystals \cite{xuSinglecrystal2021} of \ce{HfO2}   demonstrated that resolving the cationic columns alone appears to be sufficient to identify $90^\circ$ ferroelectric domains in \ce{HfO2}-based ferroelectric materials. However, the explicit atomic structure at $90^\circ$ domain walls has not be clearly resolved yet.  Compared to $90^\circ$ domain walls, $180^\circ$ domain walls in \ce{HfO2} and \ce{ZrO2} remain even much less explored.  An antipolar orthorhombic $Pbca$ phase $O_\mathrm{i}$ type \cite{leeOi2020b,ohtakaOi1995,ohtakaOi990} and a \textit{T} phase type \cite{t_type_w_2021} $180^\circ$ domain walls have recently been conjectured, but these types of domains remain to be confirmed experimentally by atomic resolution TEM imaging. An antipolar orthorhombic $Pbca$ phase denoted as $O_\mathrm{AFE}$ that differs from the $O_\mathrm{i}$ structure was observed in twinned \ce{HfO2} nanocrystals using the NCSI technique \cite{duMatter2021}, revealing $180^\circ$ domains with width of a single unit cell. To indicate the correlation between the $O_\mathrm{AFE}$ and $O_\mathrm{i}$ structures that they possess the same \textit{Pbca} space group, in this work we re-designate the former as $O_\mathrm{i^\prime}$. A question remains whether a $O_\mathrm{i^\prime}$ type domain wall exists for $180^\circ$ domains with width of more than one unit cell.

Here, based on the antipolar $O_\mathrm{i}$ and $O_\mathrm{i^\prime}$ structures,  we report two series of structures that represent two types of $180^\circ$ ferroelectric  domain walls in \ce{HfO2} and \ce{ZrO2}. We model the domain structures with different width by density functional theory (DFT) calculations and discuss the results in context of experimental observations by transmission electron microscopy. We show how the directions of polarization of 180$^\circ$ ferroelectric domains that are separated by an $O_\mathrm{i^\prime}$ type wall can be determined from experimentally observed Hf atoms even though the knowledge about the positions of O atoms is missing.

\section{methods}

All the theoretical calculations were conducted using the Quantum ESPRESSO (QE) software package \cite{QE-2020,QE-2017,QE-2009}. Structure optimization and total energy calculations were performed using DFT with the generalized gradient approximation in the Perdew–Burke–Erzenhof (GGA-PBE) formulation \cite{perdewGGA-PBE1996} and projector augmented wave (PAW) potentials \cite{blochlPAW1994a}.
A kinetic energy cut-off of 80 Rydberg (Ry) was applied to the wavefunctions. The cut-off of the charge density was 600 Ry. The Brillouin zone was sampled with an $8\times8\times8$ shifted Monkhorst-Pack (MP) grid \cite{monkhorstMP-grid1976} for the $O_\mathrm{FE}$ and $T$ phases. For the $O_\mathrm{i,n}$ and $O_\mathrm{i^\prime,n}$ structures, the Brillouin zone was sampled with a $4\times8\times8$ shifted MP grid for $\mathrm{n}=1$ otherwise with a $2\times8\times8$ shifted MP grid. The equilibrium structures were obtained by free relaxation of all atoms in the structure to the interatomic forces value of about 0.22\;meV/\AA.

Born effective charge tensors $\bm{Z}^\ast$ for every atom in each unit cell of the $O_\mathrm{i,n}$ and $O_\mathrm{i^\prime,n}$ (n = 1 to 5) structures, as well as of the paraelectric tetragonal structure were calculated using the density-functional perturbation theory (DFPT) approach via the Green's function technique \cite{baroniGreenSfunctionApproach1987}. Phonon dispersion curves of $O_\mathrm{i}$ (i.e.$O_\mathrm{i,n}$ $\mathrm{n}=1$) and $O_\mathrm{i^\prime}$  (i.e.$O_\mathrm{i^\prime,n}$ $\mathrm{n}=1$ ) structures of \ce{HfO2} and \ce{ZrO2} along high-symmetry directions were calculated by the DFPT method \cite{baroniGreenSfunctionApproach1987}, for which the ultrasoft\cite{vanderbiltUSP1990} instead of the PAW pseudopotentials were used.  There are marginal differences between the phonon dispersion curves calculated using dynamical matrices on a $4\times4\times4$ \textbf{q}-point MP grid and those using dynamical matrices on  a $2\times2\times2$ \textbf{q}-point MP grid for the $O_\mathrm{i^\prime}$ structure of \ce{HfO2}. Therefore, the dynamical matrices were calculated on a $2\times2\times2$ \textbf{q}-point MP grid for time efficiency. The acoustic sum rule was enforced. Fourier interpolation was used to obtain complete phonon dispersion curves.

\section{results}

\begin{figure}
  	\includegraphics[width=\linewidth]{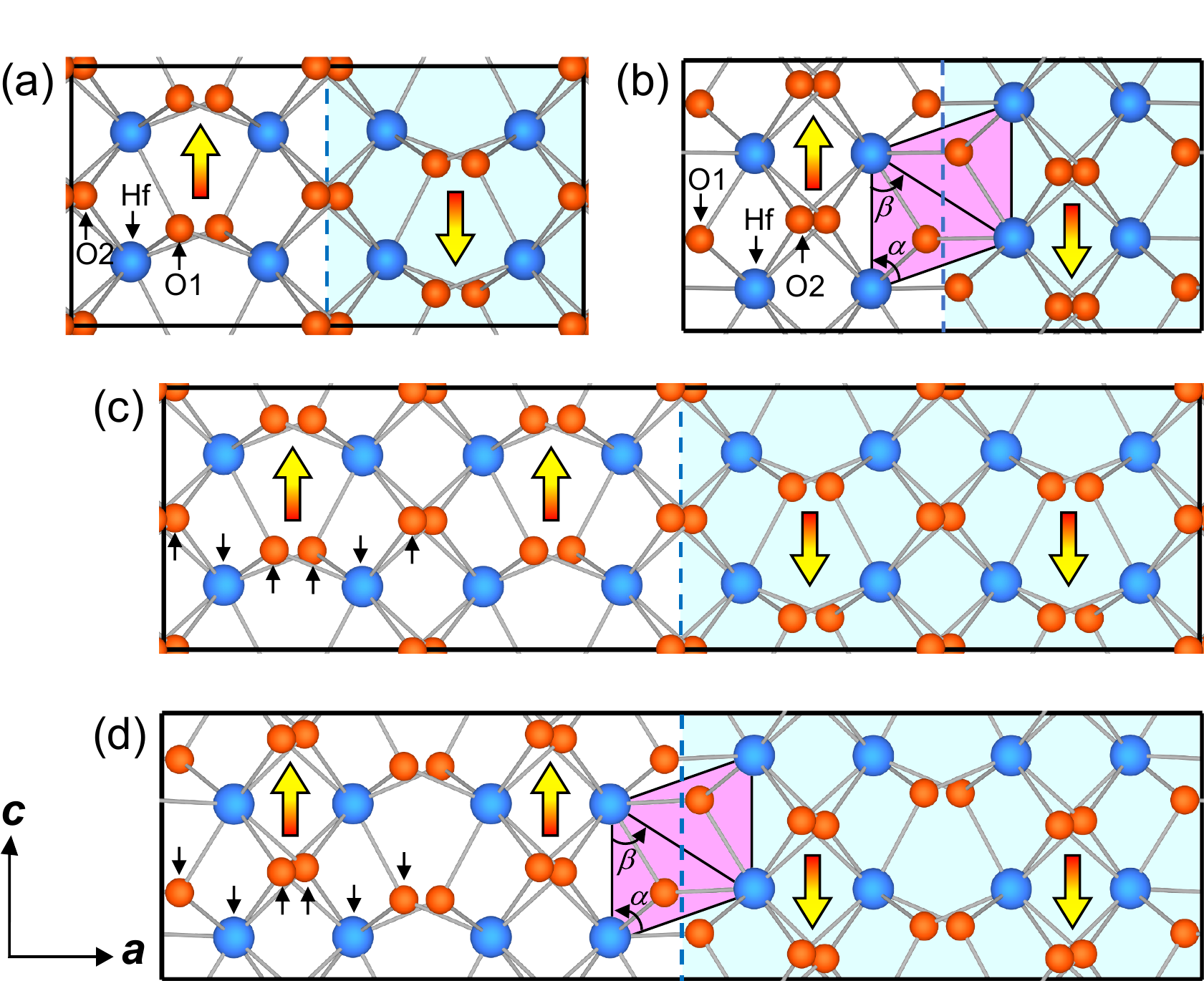}
  	\subfloat{\label{fig:str_a}}
  	\subfloat{\label{fig:str_b}}
  	\subfloat{\label{fig:str_c}}
  	\subfloat{\label{fig:str_d}}
  	\caption{Schematics of $O_\mathrm{i,n}$  (a) and (c) and $O_\mathrm{i^\prime,n}$ (b) and (d)  series of structures of \ce{MO2} (M = Hf, Zr) in $Pbca$ space group for n = 1 and 2, respectively. The black rectangle in each structure indicates the unit cell. Atoms forming the basis of the respective unit cell were indicated by short black arrows. Heads of color-filled arrows point to the direction of local polarization. Dashed lines indicated domain walls on the (200) plane. $\angle \alpha \approx 70^\circ$. $\angle \beta \approx 58^\circ$.}
  	\label{fig:str}
\end{figure}

In each structure of $O_\mathrm{i}$ and $O_\mathrm{i^\prime}$, a general position comprises one formula of \ce{MO2} (M = Zr and Hf, O1, and O2, as indicated in \cref{fig:str_a,fig:str_b}) , i.e. the basis of the structure \cite{glazerSpaceGroupsSolid2013}. These atoms forming the basis were repeated according to the symmetry of the point group of \textit{Pbca} space group into a multiplicity of 8 in the unit cell, resulting in two antiparallel $O_\mathrm{FE}$ sublattices that neighbor to each other with O2 and O1 atoms across the [200] plane (indicated by the dashed line in \cref{fig:str_a,fig:str_b}) for the $O_\mathrm{i}$ and $O_\mathrm{i^\prime}$ structure, respectively. These two structure represents 180$^\circ$ domains of the ferroelectric $O_\mathrm{FE}$ structure with width of one unit cell. 
   
In an analogous way to the $O_\mathrm{i}$ and $O_\mathrm{i^\prime}$ structures, we may put 2 formulae of \ce{MO2} of the $O_\mathrm{FE}$ structure with atoms at positions indicated by black arrows in \cref{fig:str_c,fig:str_d} as the basis in a unit cell with dimensions of $2\times1\times1$ unit cells of the $O_\mathrm{i}$ or $O_\mathrm{i^\prime}$ structure. Each of the two resulting structures comprises 180$^\circ$ domains whose width is of two unit cells of the $O_\mathrm{FE}$ structure. Indeed, it holds for any positive integer n that setting the basis of n formulae of \ce{MO2} in a $\mathrm{n}\times1\times1$ unit cells of the $O_\mathrm{i}$ or $O_\mathrm{i^\prime}$ structure results in 180$^\circ$ domains of n unit cells of the $O_\mathrm{FE}$ structure in width. This thus leads to two series of structures,  $O_\mathrm{i,n}$ and $O_\mathrm{i^\prime,n}$, in the same \textit{Pbca} space group. Lattice constants and fractional atomic coordinates of the relaxed structures with n = 1 to 5 were given in \cref{tab:hfo2_str}. 
   
We designate  the collection of the two series of structure as $O_\mathrm{AFE}$. The unit cell of each $O_\mathrm{AFE}$ structure  comprises $\mathrm{2n\times1\times1}$ $O_\mathrm{FE}$ unit cells and hence 8n formulae of \ce{MO2} (M = Hf, Zr) forming two $180^\circ$ domains. The width of each domain is of n unit cells of the $O_\mathrm{FE}$ structure. In the unit cell, the two domains were separated by a domain wall on the (200) plane as indicated by the dashed line in \cref{fig:str_a,fig:str_b,fig:str_c,fig:str_d}. Owning to the translational symmetry,  a second domain wall occurs on the (100) plane.
   
As shown in \cref{fig:str_a,fig:str_c}, when viewing along the \textit{b} axis direction the Hf atoms arrange uniformly in the two domains of opposite polarization for the $O_\mathrm{i,n}$ structures. As a result, information about the arrangement of oxygen atoms is needed in order to determine the location of the domain wall and the directions of polarization of the domains. In contrast, in the projection of the $O_\mathrm{i^\prime,n}$ structures shown in \cref{fig:str_b,fig:str_d}, the Hf atoms show a vertical offset across the domain wall (i.e., (200) plane of the \textit{Pbca} unit cell). Four neighboring Hf atoms across the domain wall form a parallelogram. The shorter diagonal divides the parallelogram into two congruent scalene triangles. The measure of the internal angle $\alpha$ is 70$^\circ$ larger than that of 58$^\circ$ for the $\beta$ [\cref{fig:str_b,fig:str_d}]. The directions of polarization of the domains can thus be determined uniquely from the arrangement of the Hf atoms alone that they are opposite to the directions of shearing of the parallelogram with respect to a rectangle.

\begin{table*}
\caption{\label{tab:hfo2_str} Lattice constants (\textit{a b c} in \AA) and fractional atomic  coordinates (\textit{x y z}) of the relaxed  $O_\mathrm{i,n}$ and $O_\mathrm{i^\prime,n}$ structures with space group \textit{Pbca}.}
\begin{ruledtabular}
\begin{tabular}{llllll}
 \multirowcell{2}[0ex][l]{n} & \multirowcell{2}[0ex][l]{items\footnote{M stands for the metal Hf and Zr atom. O1 and O2 designate oxygen atoms that are coordinated with 3 and 4 metal atoms, respectively.  The number following M, O1, and O2 indicates the index. Only atoms that constitute the basis of the respective structure are listed. The occupancy for all atoms is 1.}}& \multicolumn{2}{c}{\ce{HfO2}} & \multicolumn{2}{c}{\ce{ZrO2}}\\
 \cline{3-6}
 & & \multicolumn{1}{c}{$O_\mathrm{i,n}$} &  \multicolumn{1}{c}{$O_\mathrm{i^\prime,n}$} & \multicolumn{1}{c}{$O_\mathrm{i,n}$} &  \multicolumn{1}{c}{$O_\mathrm{i^\prime,n}$}\\
\hline
 % n = 1
\multirowcell{4}[4ex]{1}
&
\makecell[l]{
$a$ $b$ $c$\\
M\\
O1\\
O2\\}
&%HfO2 Oi,1
\makecell[l]{
9.9846 5.2096 5.0499\\% a b c in angstrom
0.1157 0.5358 0.2459\\%Hf
0.2114 0.8719 0.3757\\%O1
0.0224 0.2613 0.5020\\%O2
}
&%HfO2 Oi',1
\makecell[l]{
10.1140 5.1353 5.2827\\% a b c in angstrom
0.1377 0.4584 0.1556\\%Hf
0.0316 0.1760 0.3395\\%O1
0.2243 0.7476 0.4110\\%O2
}
&%ZrO2 Oi,1
\makecell[l]{
10.1500 5.2990 5.1320\\% a b c in angstrom
0.1157   0.5348   0.2481\\%Zr
0.2112   0.8730   0.3753\\%O1
0.0219   0.2612   0.5024\\%O2
}
&%ZrO2 Oi',1
\makecell[l]{
10.2814 5.2294 5.3641\\% a b c in angstrom
0.1380   0.4576   0.1568\\%Zr
0.03133   0.1750   0.3387\\%O1
0.2252   0.7483   0.4104\\%O2
}\\
\hline
% n= 2
\multirowcell{7}[7ex]{2}
&
\makecell[l]{$a$ $b$ $c$\\
M1\\
O11\\
O21\\
M2\\
O12\\
O22\\}
& %HfO2 Oi,2
\makecell[l]{
20.0148 5.2187 5.0472\\% a b c in angstrom
0.0578 0.5352 0.2463\\% Hf1
0.1066 0.8690 0.3802\\% O11
0.0108 0.2610 0.5026\\% O21
0.1917 0.0342 0.2410\\% Hf2
0.1430 0.3688 0.3781\\% O12
0.2401 0.7325 0.4926\\% O22
}
& %HfO2 Oi',2
\makecell[l]{
20.1130 5.2008 5.1415\\ % a b c in angstrom
0.0693 0.4615 0.1585\\ % Hf1
0.0170 0.1635 0.3260\\ % O11
0.1149 0.7624 0.4038\\ % O21
0.1837 0.9637 0.1601\\ % Hf2
0.2326 0.6320 0.2999\\ %O12
0.1372 0.2587 0.4208\\ %O22
}
& %ZrO2 Oi,2
\makecell[l]{
20.3535 5.3067 5.1293\\% a b c in angstrom
0.0578   0.5338   0.2486\\ % Zr1
0.1067   0.8697   0.3807\\ % O11
0.0105   0.2609   0.5032\\ % O21
0.1916   0.0323   0.2425\\ % Zr2
0.1430   0.3694   0.3781\\ % O12
0.2404   0.7308   0.4911\\ % O22
}
& %ZrO2 Oi',2
\makecell[l]{
20.4510 5.2916 5.2194\\ % a b c in angstrom
0.0693   0.4625   0.1602\\ % Zr1
0.0173   0.1596   0.3226\\ % O11
0.1152   0.7635   0.4034\\ % O21
0.1837   0.9653   0.1630\\ % Zr2
0.2326   0.6314   0.3013\\ % O12
0.1369   0.2600   0.4209\\ % O22
}\\
\hline
% n = 3
\multirowcell{10}[10ex]{3}
&
\makecell[l]{
$a$ $b$ $c$\\
M1\\
O11\\
O21\\
M2\\
O12\\
O22\\
M3\\
O13\\
O23\\}
& %HfO2 Oi,3
\makecell[l]{
30.0408 5.2243 5.0470\\ % a b c in angstrom
0.0385 0.5352 0.2463\\ % Hf1
0.0711 0.8687 0.3807\\ % O11
0.0072 0.2610 0.5027\\ % O21
0.1276 0.0341 0.2404\\ % Hf2
0.0952 0.3686 0.3782\\ % O12
0.1602 0.7310 0.4921\\ % O22
0.2055 0.5335 0.2412\\ % Hf3
0.2387 0.8655 0.3839\\ % O13
0.1729 0.2305 0.4923\\ % O23
}
& %HfO2 Oi',3
\makecell[l]{
30.1352 5.2161 5.1023\\ % a b c in angstrom
0.0462 0.4619 0.1589\\ % Hf1
0.0117 0.1577 0.3200\\ % O11
0.0769 0.7628 0.4051\\ % O21
0.1228 0.9646 0.1612\\ % Hf2
0.1558 0.6337 0.3032\\ % O12
0.0915 0.2600 0.4209\\ % O22
0.2113 0.4655 0.1586\\ % Hf3
0.1785 0.1335 0.3006\\ % O13
0.2437 0.7700 0.4104\\ % O23
}
&  %ZrO2 Oi,3
\makecell[l]{
30.5538 5.3117 5.1291\\ % a b c in angstrom
0.0385   0.5338   0.2486\\ % Zr1
0.0712   0.8693   0.3814\\ % O11
0.0070   0.2609   0.5033\\ % O21
0.1276   0.0319   0.2418\\ % Zr2
0.0951   0.3691   0.3783\\ % O12
0.1604   0.7292   0.4905\\ % O22
0.2056   0.5311   0.2427\\ % Zr3
0.2388   0.8657   0.3851\\ % O13
0.1727   0.2287   0.4907\\ % O23
}
&  %ZrO2 Oi',3
\makecell[l]{
30.6481 5.3049 5.1813\\ % a b c in angstrom 
0.0462   0.4632   0.1608\\ % Zr1
0.0119   0.1539   0.3168\\ % O11
0.0771   0.7639   0.4048\\ % O21
0.1228   0.9665   0.1642\\ % Zr2
0.1558   0.6333   0.3051\\ % O12
0.0913   0.2613   0.4210\\ % O22
0.2113   0.4680   0.1611\\ % Zr3
0.1784   0.1332   0.3024\\ % O13
0.2439   0.7718   0.4097\\ % O23
}\\
\hline
% n = 4
\multirowcell{13}[13ex]{4}
&
\makecell[l]{
$a$ $b$ $c$\\
M1\\
O11\\
O21\\
M2\\
O12\\
O22\\
M3\\
O13\\
O23\\
M4\\
O14\\
O24\\}
& % HfO2 Oi,4
\makecell[l]{
40.0660 5.2275 5.0469\\%a,b,c in angstrom
0.0289 0.5352 0.2463\\%Hf1 frac
0.0533 0.8686 0.3808\\% O11
0.0054 0.2611 0.5027\\%O21
0.0957 0.0341 0.2402\\%Hf2
0.0713 0.3685 0.3782\\%O12
0.1201 0.7306 0.4920\\%O22
0.1541 0.5334 0.2411\\%Hf3
0.1790 0.8651 0.3846\\%O13
0.1296 0.2301 0.4923\\%O23
0.2207 0.0333 0.2405\\%Hf4
0.1958 0.3653 0.3841\\%O14
0.2454 0.7290 0.4918\\%O24
}& \makecell[l]{
40.1588 5.2226 5.0860\\% a, b, c
0.0347   0.4620   0.1593\\ % Hf1
0.0089   0.1554   0.3176\\ % O11
0.0577  0.7629   0.4058 \\ % O21
0.0922   0.9648   0.1616\\ % Hf2
0.1169   0.6338   0.3035\\ % O12
0.0687   0.2603   0.4210\\ % O22
0.1586   0.4658   0.1587\\ % Hf3
0.1340   0.1334   0.3008\\ % O13
0.1831   0.7708   0.4108\\ % O23
0.2168   0.9663   0.1592\\ % Hf4
0.2418   0.6355   0.3043\\ % O14
0.1923   0.2710   0.4107\\ % O24
}
& %ZrO2 Oi,4
\makecell[l]{
40.7533 5.3147 5.1291\\%a b c in angstrom
0.0289   0.5338   0.2486\\%Zr1
0.0533   0.8692   0.3813\\%O11
0.0052   0.2609   0.5033\\%O21
0.0956   0.0319   0.2414\\%Zr2
0.0713   0.3691   0.3782\\%O12
0.1203   0.7289   0.4902\\%O22
0.1541   0.5310   0.2425\\%Zr3
0.1791   0.8651   0.3858\\%O13
0.1294   0.2283   0.4904\\%O23
0.2207   0.0307   0.2417\\%Zr4
0.1957   0.3653   0.3853\\%O14
0.2456   0.7271   0.4898\\%O24
}
& %ZrO2 Oi',4
\makecell[l]{
40.8465 5.3106 5.1658\\%a b c in angstrom
0.0347   0.4634   0.1612\\%Zr1
0.0091   0.1517   0.3146\\%O11
0.0579   0.7640   0.4056\\%O21
0.0922   0.9667   0.1648\\%Zr2
0.1170   0.6334   0.3058\\%O12
0.0685   0.2616   0.4213\\%O22
0.1586   0.4683   0.1614\\%Zr3
0.1339   0.1331   0.3030\\%O13
0.1833   0.7725   0.4102\\%O23
0.2168   0.9689   0.1620\\%Zr4
0.2419   0.6355   0.3069\\%O14
0.1922   0.2728   0.4101\\%O24
}\\
\hline
% n = 5
\multirowcell{16}[16ex]{5}
&
\makecell[l]{$a$ $b$ $c$\\
M1 \\
O11\\
O21\\
M2\\
O12\\
O22\\
M3\\
O13\\
O23\\
M4\\
O14\\
O24\\
M5\\
O15\\
O25\\}
& %HfO2 Oi,5
\makecell[l]{
50.0913 5.2296 5.0467\\ % a,b,c in angstrom 
0.0231   0.5352   0.2464\\ % Hf1
0.0427   0.8686   0.3808\\ % O11
0.0043   0.2611   0.5027\\ % O21
0.0765   0.0342   0.2403\\ % Hf2
0.0570   0.3685   0.3783\\ % O12
0.0961   0.7305   0.4922\\ % O22
0.1232   0.5334   0.2413\\ % Hf3
0.1432   0.8650   0.3849\\ % O13
0.1037   0.2300   0.4924\\ % O23
0.1765   0.0332   0.2406\\ % Hf4
0.1566   0.3652   0.3844\\ % O14
0.1963   0.7287   0.4920\\ % O24
0.2234   0.5332   0.2407\\ % Hf5
0.2434   0.8648   0.3851\\ % O15
0.2036   0.2287   0.4920\\ % O25
}
& %HfO2 Oi',5
\makecell[l]{
50.1831 5.2262 5.07713\\ % a,b,c in angstrom 
0.0278   0.4621   0.1595\\ % Hf1
0.0072   0.1542   0.3164\\ % O11
0.0462   0.7629   0.4062\\ % O21
0.0738   0.9648   0.1618\\ % Hf2
0.0936   0.6337   0.3035\\ % O12
0.0550   0.2604   0.4211\\ % O22
0.1270   0.4659   0.1588\\ % Hf3
0.1072   0.1333   0.3008\\ % O13
0.1466   0.7709   0.4108\\ % O23
0.1736   0.9664   0.1594\\ % Hf4
0.1935   0.6356   0.3045\\ % O14
0.1539   0.2712   0.4107\\ % O24
0.2267   0.4666   0.1590\\ % Hf5
0.2067   0.1355   0.3043\\ % O15
0.2464   0.7717   0.4106\\ % O25
}
& %ZrO2 Oi,5
\makecell[l]{
50.9530 5.3165 5.1291\\ % a,b,c in angstrom
0.0231   0.5338   0.2486\\%Zr1
0.0427   0.8693   0.3814\\%O11
0.0042   0.2609   0.5033\\%O21
0.0765   0.0319   0.2415\\%Zr2
0.0570   0.3691   0.3782\\%O12
0.0962   0.7288   0.4903\\%O22
0.1232   0.5310   0.2426\\%Zr3
0.1433   0.8650   0.3860\\%O13
0.1035   0.2282   0.4905\\%O23
0.1765   0.0306   0.2417\\%Zr4
0.1565   0.3652   0.3854\\%O14
0.1964   0.7269   0.4898\\%O24
0.2234   0.5305   0.2418\\%Zr5
0.2434   0.8647   0.3863\\%O15
0.2035   0.2268   0.4898\\%O25
}
& %ZrO2 Oi',5
\makecell[l]{
51.0456 5.3137 5.1575\\ % a b c in angstrom
0.0278   0.4634   0.1613\\ % Zr1
0.0073   0.1507   0.3135\\ % O11
0.0463   0.7640   0.4059\\ % O21
0.0738   0.9668   0.1649\\ % Zr2
0.0936   0.6333   0.3056\\ % O12
0.0548   0.2617   0.4212\\ % O22
0.1269   0.4684   0.1613\\ % Zr3
0.1072   0.1330   0.3027\\ % O13
0.1467   0.7726   0.4101\\ % O23
0.1736   0.9691   0.1620\\ % Zr4
0.1936   0.6357   0.3071\\ % O14
0.1538   0.2729   0.4100\\ % O24
0.2266   0.4693   0.1615\\ % Zr5
0.2066   0.1355   0.3068\\ % O15
0.2465   0.7736   0.4098\\ % O25 
}\\
\end{tabular}
\end{ruledtabular}
\end{table*}

\begin{figure}
	\includegraphics[width=\linewidth]{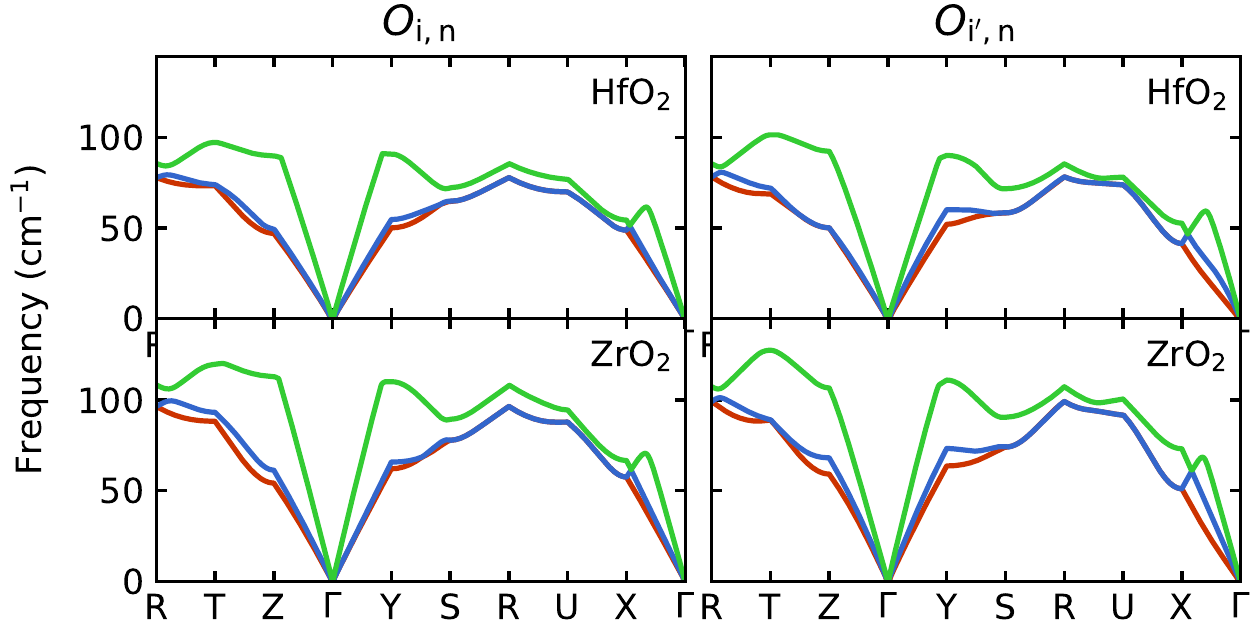}
	\caption{Acoustic phonon dispersion curves of  $O_\mathrm{i}$ and $O_\mathrm{i^\prime}$, i.e. $O_\mathrm{i,n}$ and $O_\mathrm{i^\prime,n}$ for n = 1, structures of \ce{HfO2} and \ce{ZrO2}  along high-symmetry directions.}
	\label{fig:phonon}
\end{figure}

\Cref{fig:phonon} shows acoustic phonon dispersion curves of $O_\mathrm{i}$ and $O_\mathrm{i^\prime}$  structures of \ce{HfO2} and \ce{ZrO2} along high-symmetry directions. For all the structures, no imaginary or negative frequencies were observed along the high symmetry directions in the Brillouin zone with enforcing the acoustic sum rule. This confirms the stability of these structures in terms of lattice dynamics.  The dispersion curves appear to be remarkably similar between \ce{HfO2} and \ce{ZrO2} and between the $O_\mathrm{i}$ and $O_\mathrm{i^\prime}$ structures. The dispersion curves of \ce{ZrO2} show slightly higher frequencies compared to those of \ce{HfO2} for the same type of structure at high-symmetry points other than the $\Gamma$ point. This is explained by the lighter atomic mass of Zr than that of the Hf. The result for the $O_\mathrm{i}$ structure of \ce{HfO2} is in well agreement with the calculations by a different implementation of the DFPT using ABINIT package \cite{zhou_HfO2_phonon_2014}.

\begin{figure}
	\includegraphics[width=\linewidth]{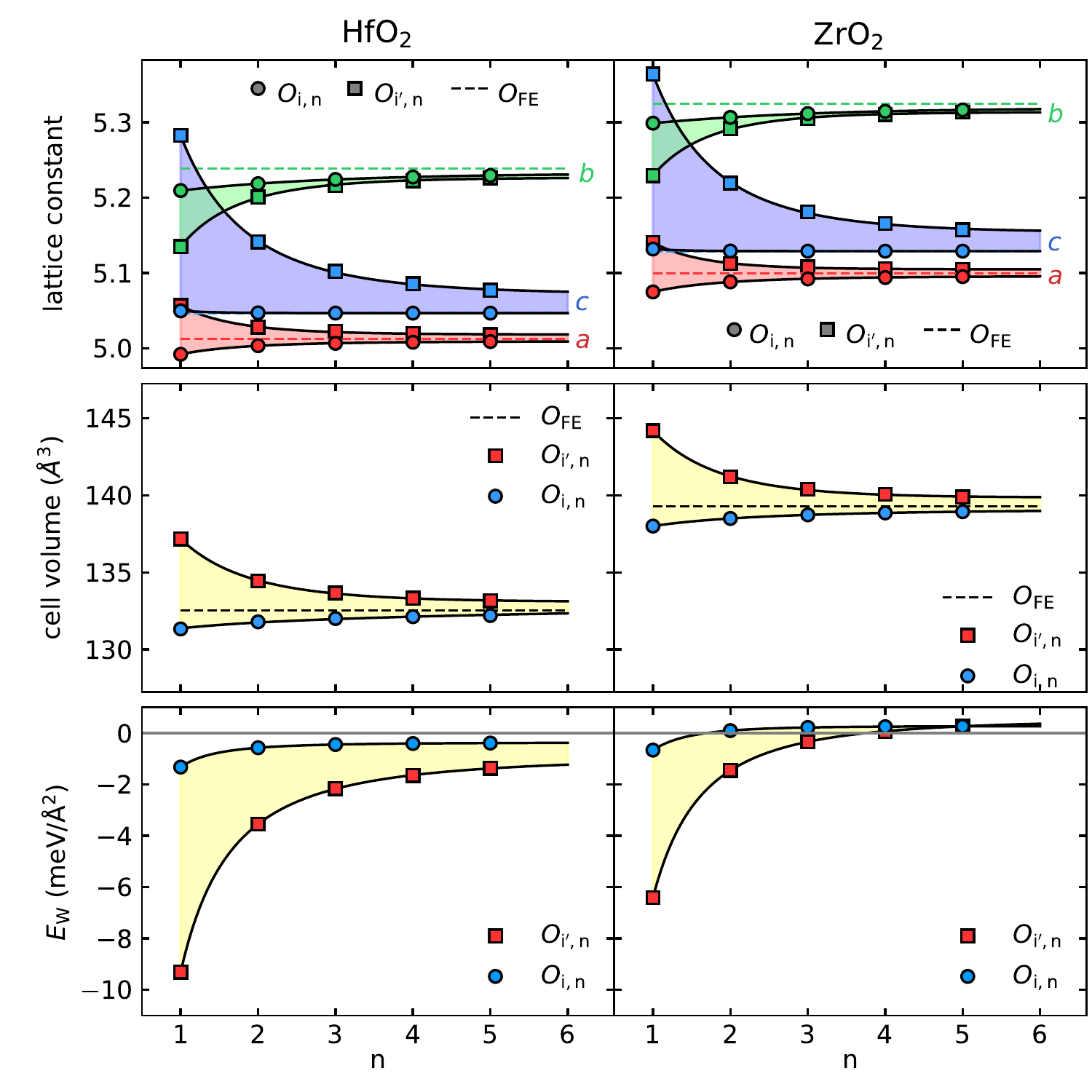}
	\caption{Lattice constant, cell volume, and domain wall energy ($E_\mathrm{W}$) depending on the domain width n, i.e., the number of unit cells of $O_\mathrm{FE}$ of the same direction of polarization in an unit cell of $O_\mathrm{i,n}$ and $O_\mathrm{i^\prime,n}$ structure. The lattice constant \textit{a} and cell volume are normalized to per unit cell of $O_\mathrm{FE}$.}
	\label{fig:celpar_edw}
\end{figure}

Lattice constant, cell volume, and domain wall energy ($E_\mathrm{W}$) of the $O_\mathrm{i,n}$ and $O_\mathrm{i^\prime,n}$ structures of \ce{HfO2} and \ce{ZrO2} as a function of the domain width n were calculated using density functional theory (DFT). As shown in \cref{fig:celpar_edw}, these properties show a remarkable similarity between \ce{HfO2} and \ce{ZrO2} but follow different trends between the $O_\mathrm{i,n}$ and $O_\mathrm{i^\prime,n}$ structures. Properties of the $O_\mathrm{i^\prime,n}$ structures show more pronounce dependence on n than those of the $O_\mathrm{i,n}$ structures. For the $O_\mathrm{i,n}$ structures, the lattice constants \textit{a} and \textit{b} increase but \textit{c} remains a constant value of about the same as that of the $O_\mathrm{FE}$ structure with the increase of n. In contrast, for the $O_\mathrm{i\prime,n}$ structures, the lattice constants \textit{a} and \textit{c} decrease but \textit{b} increases with the increase of n. 
The cell volume, normalized to per unit cell of $O_\mathrm{FE}$, of the $O_\mathrm{i,n}$ structures increases while that of the $O_\mathrm{i\prime,n}$ structures decreases, and both are approaching that of the $O_\mathrm{FE}$ structure, with the increase of n.

The domain wall energy per area ($E_\mathrm{w}$) was calculated from the total energy of unit cell of the $O_\mathrm{AFE}$, i.e. $O_\mathrm{i,n}$ and $O_\mathrm{i^\prime,n}$, and $O_\mathrm{FE}$ structures according to \cref{eq:edw}.

\begin{equation}
E_\mathrm{W} = \frac{E_{O_\mathrm{AFE}} - 2\mathrm{n}E_{O_\mathrm{FE}}}{2A} \label{eq:edw}
\end{equation}
Here $A=b\cdot c$ stands for the area per unit cell of an individual domain wall on the (100) or (200)  plane of the $O_\mathrm{AFE}$ structure. 
The $b$ and $c$ are the two lattice constants. 

The calculated values of domain wall energy are shown in \cref{fig:celpar_edw}.
For both \ce{HfO2} and \ce{ZrO2}, the wall energy increases with domain width n at a decreasing rate, irrespective of $O_\mathrm{i,n}$ and $O_\mathrm{i^\prime,n}$ series of structures. Given the same domain width n, the $O_\mathrm{i,n}$ series generally shows a higher wall energy than that of the $O_\mathrm{i^\prime,n}$. This difference becomes smaller when domain width n is larger because the increase rate of the wall energy of the former  decreases much more rapidly than that of the latter. In particular, for \ce{ZrO2} at n = 4 and 5 the respective wall energy of $O_\mathrm{i,n}$ and $O_\mathrm{i^\prime,n}$ is at about the same value very close to zero.

\begin{figure*}
	\includegraphics[width=\linewidth]{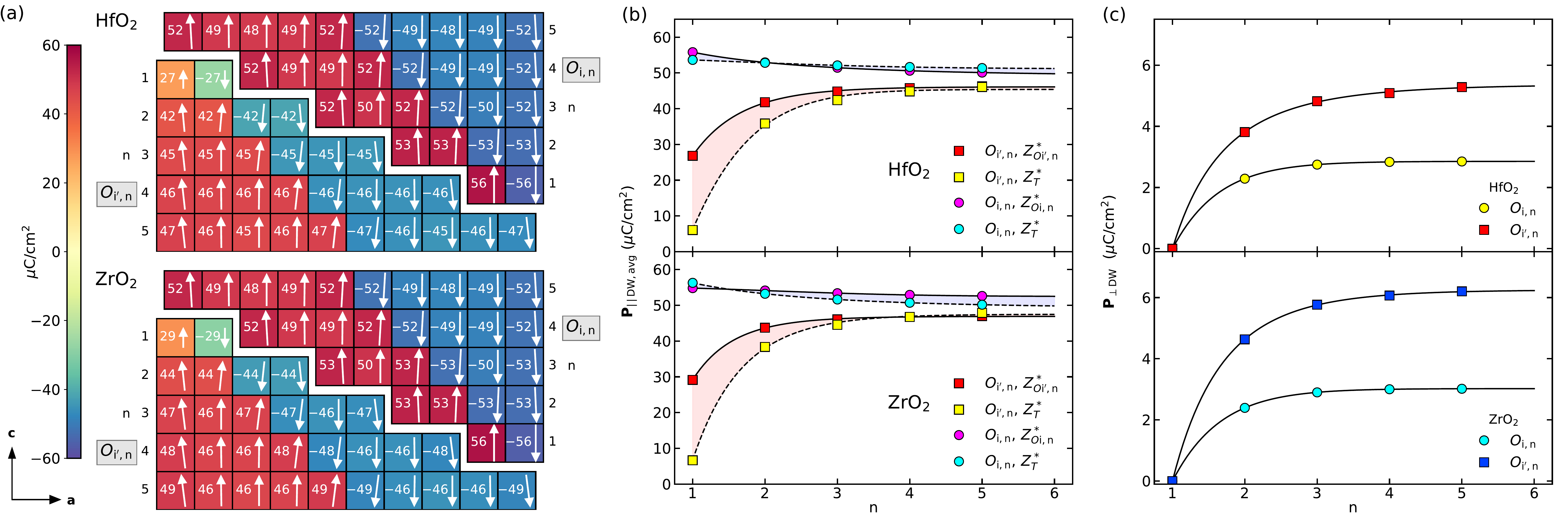}
	\subfloat{\label{fig:p_a}}
	\subfloat{\label{fig:p_b}}
	\subfloat{\label{fig:p_c}}
	\caption{Polarization depending on the domain width n of the  $O_\mathrm{i,n}$ and $O_\mathrm{i^\prime,n}$ series of structures of \ce{HfO2} and \ce{ZrO2}. (a) Polarization vectors are projected on the \textit{a}-\textit{c} plane and are represented as arrows. The polarization was calculated using the Born effective charge tensors $\bm{Z}^\ast$ of the corresponding structure. The arrow head points to the directions of the polarization in each local $O_\mathrm{FE}$ unit cell indicated by the square. The number next to each arrow indicates the magnitude of the polarization, which is positive when it points to the $\bm{c}$-axis direction.  (b) Magnitude of polarization component parallel to the domain wall ($\bm{P}_{\parallel\mathrm{W}}$) averaged over the domain. The polarization was calculated using the Born effective charge tensors of  the corresponding $O_\mathrm{AFE}$ structure ($\bm{Z}^*_{O\mathrm{i,n}}$ and $\bm{Z}^*_{O\mathrm{i^\prime,n}}$) and using those of the tetragonal structure ($\bm{Z}^*_{T}$). (c) Polarization component pointing perpendicularly to the domain wall ($\bm{P}_{\perp\mathrm{W}}$) for locations next to the domain wall calculated using the $\bm{Z}^\ast$ of the corresponding structure.} 
	\label{fig:p}
\end{figure*}

Born effective charge tensors $\bm{Z}^\ast$ for every atom in the unit cell of each  $O_\mathrm{AFE}$ (i.e. $O_\mathrm{i,n}$ and $O_\mathrm{i^\prime,n}$) structure, as well as of the paraelectric tetragonal structure were calculated using the DFPT approach via the Green's function technique \cite{baroniGreenSfunctionApproach1987}. The local polarization $\bm{P}$ in direction $\beta$ per  $O_\mathrm{FE}$ unit cell was thereafter calculated from the displacement $\bm{u}_{i,\alpha}$ of the i-th atom in direction $\alpha$ with respect to the tetragonal structure according to \cref{eq:p}.
\begin{equation}
\bm{P}_{\beta} = \frac{e}{V_\mathrm{cell}}
 \sum\limits_{\mathrm{i}}^{} \sum\limits_{\alpha}^{}   \bm{Z}^{*}_{\mathrm{i},\alpha\beta} \bm{u}_{\mathrm{i},\alpha}
 \label{eq:p}
\end{equation}
Here $V_\mathrm{cell}$ is the volume of the local $O_\mathrm{FE}$ unit cell, $\bm{Z}^{*}_{\mathrm{i},\alpha\beta}$ is the Born effective charge for i-th atom.  and the summations run over all the atoms and the Cartesian directions.

Local polarization projected on the \textit{a-c} plane was shown as arrows in \cref{fig:p_a}, where the arrow head points to the polarization direction and the arrow length represent the magnitude of polarization. The prominent component of the local polarization is parallel to the domain wall [\cref{fig:p_a,fig:p_b}].  For  both \ce{HfO2} and \ce{ZrO2},  the  polarization of $180^\circ$ domains of the $O_\mathrm{i,n}$ series of structures exhibits a larger value than that of $180^\circ$ domains of the $O_\mathrm{i^\prime,n}$ series of structures at the same n. This difference is more notable  when n is small. The former progressively decreases, but the latter more abruptly increases, and both approach about 50 $\mu\mathrm{C/cm^2}$ with the increase of n.  $O_\mathrm{i,n}$ and $O_\mathrm{i^\prime,n}$ series of structures of \ce{ZrO2} behave remarkably  similarly to those of \ce{HfO2}  in terms of both the values of magnitude of polarization and the manner of its dependence on n.  The results reveal marginal dependence of the magnitude of local polarization along the $\bm{c}$-axis direction on the distance of the location to the $180^\circ$ domain wall irrespective of \ce{HfO2} and \ce{ZrO2} structures [\cref{fig:p_a}]. Except for n = 1, the local polarization at locations that are next to the domain wall exhibits a noticeable component $\bm{P}_{\perp W}$ pointing perpendicularly to the domain wall [\cref{fig:p_a,fig:p_c}]. This suggests that these domain walls appear to be positively charged.  The magnitude of the $\bm{P}_{\perp W}$ is larger for the $O_\mathrm{i^\prime,n}$ than that of the $O_\mathrm{i,n}$  series of structures and increases with the increase of n for both the series of structures \cref{fig:p_c}. 

\section{discussion}

It was assumed that observation of both Hf and O atoms are necessary in order to resolve 180$^\circ$ domains \cite{grimleyChTEM2019}. This appears true for domains that are separated by an $O_\mathrm{i}$ type wall because of the coherent arrangement of Hf atoms across the wall. However, it is worthwhile to point out that for domains that are separated by an $O_\mathrm{i^\prime}$ type wall the polarization directions can be uniquely determined from the observation of Hf atoms even the knowledge about the O atoms is missing. 

\begin{figure}
	\begin{center}
	\includegraphics[width=0.875\linewidth]{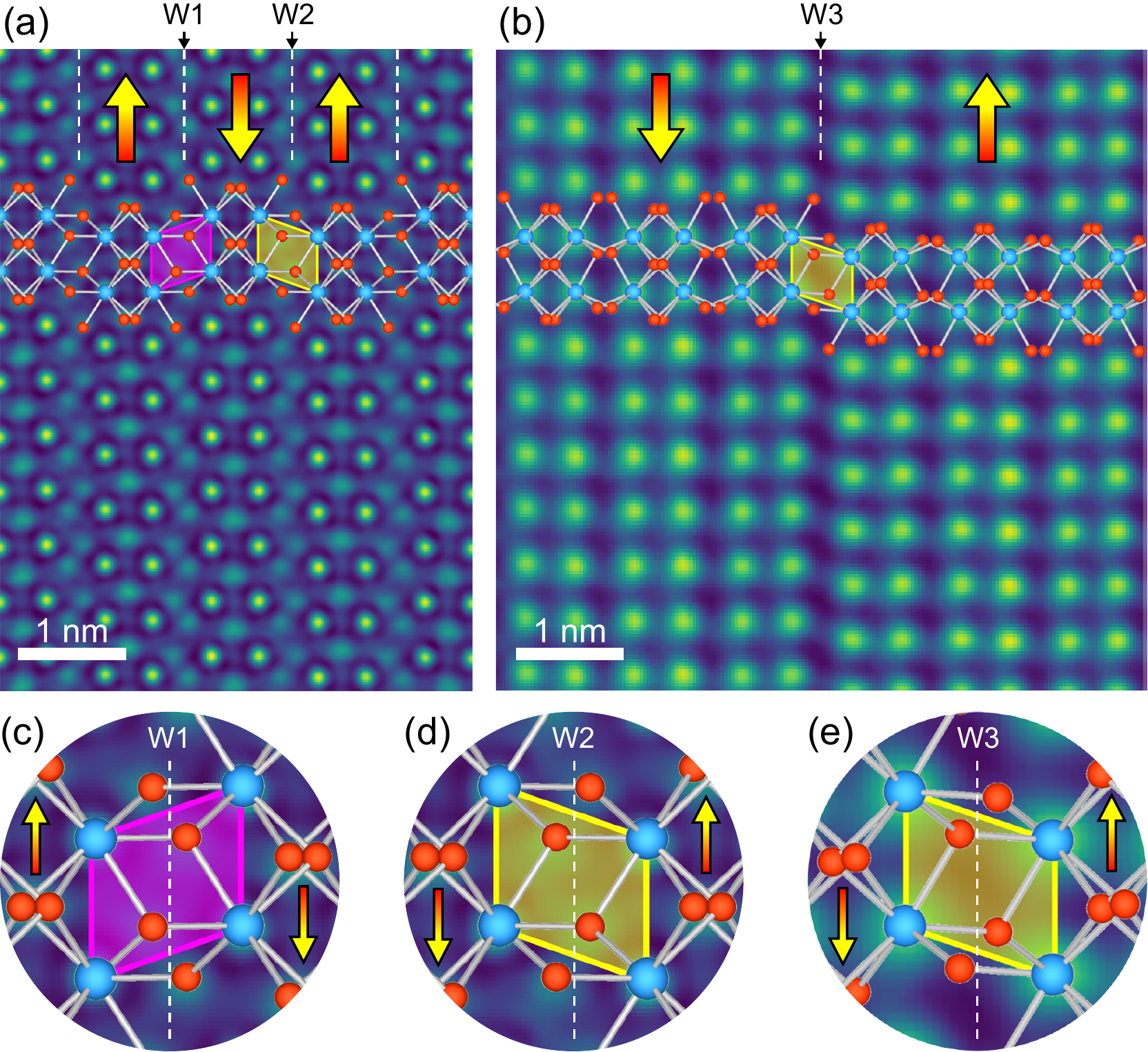}
	\end{center}
	\subfloat{\label{fig:tem_a}}
	\subfloat{\label{fig:tem_b}}
	\subfloat{\label{fig:tem_c}}
	\subfloat{\label{fig:tem_d}}
	\subfloat{\label{fig:tem_e}}
	\caption{$O_\mathrm{i^\prime}$ structural  type $180^\circ$ domain walls experimentally observed  in (a) NCSI CTEM  and (b) HAADF STEM  images of a \ce{HfO2} nanocrystal and film with zone axis of <010>, respectively. Domain walls were labeled as W1 to W3. The images in (a) and (b) are produced by rotation and averaging in the direction parallel to the domain walls of the Fig. 2(a) of Ref. \cite{duMatter2021} and about the region R1 of Fig. 5(b) of Ref. \cite{grimleyDomain2018} , respectively.  (c) to (e) Atomic structural details of the domain walls. Atomic structural schematics (Hf: blue, O: red, and bond: gray) were superposed in the images. The heads of color filled arrows indicate the directions of polarization.}
	\label{fig:tem}
\end{figure}

The $180^\circ$ domains of the  $O_\mathrm{i^\prime}$ structural type with domain width of one unit cell of $O_\mathrm{FE}$ were experimentally observed from colloidal \ce{HfO2} nanocrystals using the negative spherical aberration ($\mathrm{C_S}$) imaging (NCSI)  technique \cite{duMatter2021}. As shown in  \cref{fig:tem_a}, two domain walls separating the domains with polarization of $\uparrow\downarrow$ and $\downarrow\uparrow$ configurations were indicated as W1 and W2. The details of the atomic structure of the domain wall were shown in \cref{fig:tem_c,fig:tem_d}. The configurations of Hf atoms adjacent to the domain walls were highlighted with parallelograms. The directions of polarization of the domains W1 and W2 were unambiguously determined as a result of the observation of both Hf and O atoms in the NCSI CTEM image. The polarization directions are opposite to the directions of shearing of the parallelogram with respect to a rectangle. This observed correlation between the Hf configuration  and  the direction of polarization confirms that for the  $O_\mathrm{i^\prime}$ structural type  $180^\circ$ domains the direction of polarization in the domains can be determined uniquely from the arrangement of Hf atoms even without the knowledge about the O atom. This helps to interpret the experimental results obtained using HAADF STEM by which only Hf atoms can be resolved.

An HAADF image of a \ce{HfO2} thin film shown in \cref{fig:tem_b}, which was obtained by averaging around the region R1 of the image in Fig. 5(b) of Ref. \cite{grimleyDomain2018} along the direction parallel to the domain wall labeled as W3. In the original report, the domain wall was interpreted as a `twin boundary' due to the missing information about the O atoms. In fact, the observed structure is in a well agreement with $180^\circ$ domains of the  $O_\mathrm{i^\prime}$ structural type with domain width of several unit cells of the $O_\mathrm{FE}$ structure. The arrangement of Hf atoms adjacent to the domain wall W3 is remarkably similar to that of W2. The polarization directions of the domains were inferred from their correlation to the Hf configuration as discussed above, which formed a $\downarrow\uparrow$ configuration.

The experimental observations of the $O_\mathrm{i^\prime}$ structural type $180^\circ$ domain walls in both colloidal \ce{HfO2} nanocrystals and thin films possibly imply that the $O_\mathrm{i^\prime}$ structural type $180^\circ$ domain walls is more favorable than those of the  $O_\mathrm{i}$ type in reality. This can be understood by that the formation of the former will reduce more energy of the system than that of the latter [\cref{fig:celpar_edw}]. Experimental evidence of the  $O_\mathrm{i}$ type $180^\circ$ domain wall has not been reported so far. Such evidence requires observation of O atoms as Hf atoms are arranged coherently across the $O_\mathrm{i}$ type $180^\circ$ domain wall [\cref{fig:str_a,fig:str_c}].

{It appears to be of interest to develop techniques to growth 180$^\circ$ domains of given wall types and width in \ce{HfO2} and \ce{ZrO2} in a controllable way. Based on these progresses, the dependence of polarization on the domain type and width may be experimentally observed using atomic resolution TEM imaging techniques that allow resolving both Hf and O atoms, thereby making it possible to compare these experimental observations with our theoretical calculations. Moreover future studies are required to clarify how the two different types of 180$^\circ$ domains play a role in the polarization switching.}

\section{conclusion}
In conclusion, we have proposed two series of structures that not only allow systematic modeling of $180^\circ$ ferroelectric domains in \ce{HfO2} and \ce{ZrO2}{, but also provide more thorough interpretation and hence deeper understanding of experimental observations}. Our calculations have shown that the properties of the $180^\circ$ domains with walls of the  $O_\mathrm{i^\prime}$ type have significantly stronger dependence on the domain width than those of the $O_\mathrm{i}$ type. We have demonstrated that the observation of the cationic atoms alone suffices to to determine the directions of polarization for the $180^\circ$ domains separated by the $O_\mathrm{i^\prime}$ type walls and have shown how to do it without the knowledge about positions of O atoms. Our {results provide} a more thorough interpretation of experimental observations of domains by the {widely} used HAADF STEM technique and {are expected to} help further systematic theoretical studies
on the $180^\circ$ ferroelectric domains and ferroelectric switching properties in \ce{HfO2} based ferroelectric materials.

%\nocite{*}
\begin{acknowledgments}
We acknowledge support from German Research Foundation (DFG) under the Grant SFB917 Nanoswitches. DFT simulations were performed with computing resources granted by RWTH Aachen University under projects rwth0646 and rwth0786.% We thank Joachim Mayer and Rafal E. Dunin-Borkowski for their support.
\end{acknowledgments}

\bibliography{dw180}% Produces the bibliography via BibTeX.

\end{document}